\documentclass{ws-ijmpa}

\usepackage{latexsym}
\usepackage{graphicx}
\usepackage{float}
\usepackage{amsmath}

\usepackage{amsfonts}
\newcommand{\beq}{\begin{equation}} \newcommand{\eeq}{\end{equation}}
\newcommand{\beqa}{\begin{eqnarray}} \newcommand{\eeqa}{\end{eqnarray}}

\newcommand{\mms}[2]{\ensuremath{\tilde m^2_{#1 #2}\!~}}
\newcommand{\ms}[2]{\ensuremath{\tilde m_{#1 #2}\!~}}
\def\xst1{\ensuremath{x_{\tilde t_1}}}

\begin{document}

\markboth{A.V.~Bednyakov, D.I.~Kazakov and \c{S}.H.~Tany\i ld\i z\i}
{SUSY Enhancement  of Heavy Higgs Production}

%
%

\title{SUSY ENHANCEMENT OF HEAVY HIGGS PRODUCTION}

\author{A.~V.~BEDNYAKOV} 

\address{Bogoliubov Laboratory of Theoretical Physics, Joint
Institute for Nuclear Research, \\
Dubna, Moscow Region, 114980, Russia,\\
bednya@theor.jinr.ru}

\author{D.~I.~KAZAKOV}

\address{
Institute for Theoretical and Experimental Physics, 
Moscow, 117218,
Russia; \\
Bogoliubov Laboratory of Theoretical Physics, Joint
Institute for Nuclear Research, \\
Dubna, Moscow Region, 114980, Russia,\\
kazakovd@theor.jinr.ru
}

\author{\c{S}.~H.~TANYILDIZI}

\address{Bogoliubov Laboratory of Theoretical Physics, Joint
Institute for Nuclear Research, \\
Dubna, Moscow Region, 114980, Russia,\\
hanif@theor.jinr.ru
}

\maketitle


\begin{abstract}
We study the cross-section of heavy Higgs production at the LHC within the framework of the Constrained MSSM. 
It is not only enhanced by $\tan^2\beta$ but sometimes is also enhanced by the squark contribution.  First, we consider the universal scenario within mSUGRA and find out that to get the desired enhancement one needs large negative values of $A_0$, which seems to be incompatible with the $b\to s\gamma$ decay rate. To improve the situation, we release the unification requirement in the Higgs sector. Then 
it becomes possible to satisfy all requirements simultaneously and enhance the squark contribution. 
The latter can gain a factor of several units  increasing the overall cross-section  which, however, is still
smaller than the cross-section of the associated $Hb\bar b$ production.
We consider also some other consequences of the chosen benchmark point.
\end{abstract}

\ccode{PACS numbers: 12.60.Jv, 14.80.Da}

\section{Introduction}
\label{sec:intro}

With the launch of the LHC the expectations for discovery of the Higgs boson and possible new physics became actual.
As usual, the production of heavy particles is suppressed by their masses, so that one expects to find the light particles
first. However, sometimes the heavy particle production can be enhanced by some factors. This is exactly what happens with
the heavy Higgs production in the MSSM\cite{Djouadi:2005gj,Dittmaier:2011ti}. We study this enhancement for the case of gluon fusion and show that
not only the $\tan\beta$ enhancement takes place\cite{Beskidt:2010va}, 
but also there is an additional source of enhancement due to the squark contribution in the loops.

It is well known that the Higgs boson production at hadron colliders within the SM mainly goes 
through the gluon fusion process\cite{Georgi:1977gs}. It is 
the triangle loop diagram (see Fig.~\ref{fig:higgsprod}) that gives the main contribution. This is also true in the MSSM, though in this case the associated production with two b-quarks (two b-jets) is even more 
favorable\cite{Djouadi:2005gj}. The latter process is realized at the tree level and, hence, has no new virtual particles involved contrary to the loop diagrams. 
Nevertheless, the triangle diagrams do not give additional b-jets in the final states and presumably can be distinguished from the associated production by b-tagging of these jets. 

Since we are talking about new particles in the loop, their contribution depends on their masses. The smaller the mass, the bigger is the contribution. At the same time, the squark contribution is also proportional to the quark mass, so only the third generation essentially plays any role.  For numerical analysis we need the values of  squark masses and mixings. We proceed in two ways: First, we consider  the usual MSSM universal high energy parameters ($m_0,m_{1/2},A_0$, and $\tan\beta$), evaluate masses and mixings, and calculate the cross-section for various points of parameter space. We find the areas in the parameter space where the loop enhancement takes place. This requires light top-squarks which  is possible for very large and negative values of $A_t$ that implies negative $A_0$. Then we consider the fulfillment of  various constraints such as 
$B\to X_s \gamma$,\cite{Barberio:2008fa} $B_s\to \mu^+\mu^-$,\cite{Bmumuexp,Bmumuexp2} $g-2$ of muon (see, e.g., Ref.~\refcite{g-2}), 
relic density of the Dark Matter (DM)\cite{Komatsu:2008hk}, electroweak precision data on $M_W$ and 
$\sin^2\theta_{eff}$ (see, e.g., Ref.~\refcite{Heinemeyer:2004gx}), and Higgs and superpartner searches in this region. We find out that the considered universal scenarios with large negative $A_0$ are not compatible with the $b\to s\gamma$ constraint. To avoid this problem and to have the cross-section at the level of a few pb, we release the universality constraint and allow the non-universal Higgs mass (NUHM) terms\cite{Ellis:2002wv}. 
As independent variables we take the Higgs mixing term $\mu$ and the CP-odd heavy Higgs boson mass $m_A$. 

Our overall conclusion is that for a relatively light stop and moderately heavy $H_0$ one can reach essential enhancement of heavy Higgs production albeit  in the restricted region of the parameter space.  Simultaneously, one gets relatively high cross-section for the stop pair-production which might also be of interest in view of SUSY searches.

\section{Cross-section for Heavy Higgs Production in the MSSM} 
\label{sec:production}
\begin{figure}\vspace{-0.5cm}
\begin{center}
\includegraphics[width=12cm]{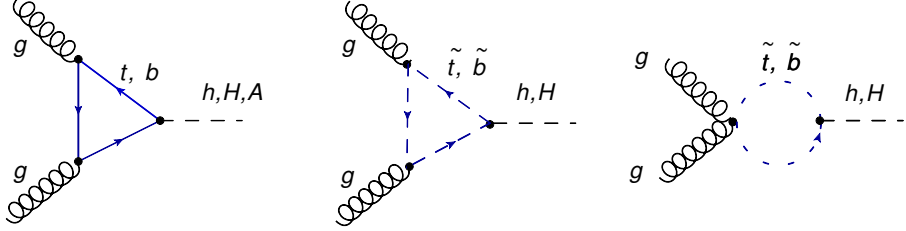}
\caption{The leading order (LO) diagrams for the Higgs boson production via gluon fusion.\label{fig:higgsprod} }
\end{center}
\end{figure}

The Feynman diagrams describing the Higgs production via gluon fusion are shown in Fig.~\ref{fig:higgsprod},
where the last ones are due to squarks in the intermediate states. As was stated earlier, all the contributions are proportional to the quark masses, so only the third generation is relevant.

 The cross-section for the Higgs boson production  with account of gluon distribution functions  is given 
 by\cite{Djouadi:2005gj,Georgi:1977gs,Spira:1993bb}
 \beq \sigma_{Higgs} = \frac{1}{32}\int_0^1 dx_1 dx_2\  g[x_1]\  g[x_2]\  |{\cal M}_{Higgs}|^2 \frac{ 2\pi}{m_{Higgs}^2} \delta(
   E^2 x_1 x_2 - m_{Higgs}^2), \label{eq:sigma} \eeq
  where $g[x]$ is the gluon distribution function inside the proton that implicitly depends on the factorization scale $Q$. 
  In our case, we take it equal to the Higgs boson mass.

The matrix elements corresponding to the above diagrams are (we use the notation where $v=175$  GeV)\cite{haber}
 \begin{eqnarray}
{\cal M}_{h}&=&\frac{\alpha_s}{4\pi}  \frac{m_h^2}{2\sqrt{2}v}
   \left(\frac{\cos\alpha}{\sin\beta} F_{1/2}^h[\frac{4m_t^2}{m_h^2}]  - \frac{\sin\alpha}{\cos\beta}F_{1/2}^h[\frac{4m_b^2}{m_h^2}] \right),
   \nonumber\\
{\cal M}_{H}&=&\frac{\alpha_s}{4\pi}  \frac{m_H^2}{2\sqrt{2}v}
   \left(\frac{\sin\alpha}{\sin\beta}F_{1/2}^H[\frac{4m_t^2}{m_H^2}] +\frac{\cos\alpha}{\cos\beta}F_{1/2}^H[\frac{4m_b^2}{m_H^2}] \right),
   \label{eq:quark}\\
{\cal M}_{A}&=&\frac{\alpha_s}{4\pi}  \frac{m_A^2}{2\sqrt{2}v}
   \left( \frac{\cos\beta}{\sin\beta}F_{1/2}^A[\frac{4m_t^2}{m_A^2}] +\frac{\sin\beta}{\cos\beta} F_{1/2}^A[\frac{4m_b^2}{m_A^2}]\right), \nonumber
\end{eqnarray}  
where the angle $\alpha$ is the neutral Higgs mixing angle defined by $\tan2\alpha=\frac{m_A^2+M_Z^2}{m_A^2-M_Z^2}\tan2\beta$ and is typically equal to $\alpha\approx \beta -\pi/2$.

It should be noted that by definition $\alpha\in [-\pi/2,0]$, so that $\sin \alpha < 0$, and the sign of the 
t-quark contribution
is different for the light $h$ and heavy $H$ Higgs boson matrix elements.  
It is known\cite{Djouadi:2005gj} that for the lightest Higgs boson $h$ (with the mass $m_h < 400-500~$ GeV) 
the loops with the bottom and top quarks interfere destructively. In contrast, in the case of the heavy boson $H$ 
the interference is constructive and becomes destructive only when the mass of the heavy boson is above 400-500 GeV.

As one can see from Eq.~(\ref{eq:quark}), the light Higgs boson $h$ production is almost not influenced by $\tan\beta$, while for heavy Higgses $H$ and $A$ the contribution of the b-quark is enhanced by $\tan\beta$ and that of the t-quark is suppressed by $\tan\beta$.  Hence, for high $\tan\beta$ (which is of interest for us due to the enhancement of the cross-section) only the b-quark is essential.

The addition of squarks is achieved by the following modification\footnote{We have corrected some misprints in Ref.~\refcite{haber}.}:
\begin{subequations}
 \begin{eqnarray}
\!\!\!\!\!\Delta{\cal M}_{h}&=&\frac{\alpha_s}{4\pi}  \frac{m_h^2}{2\sqrt{2}v}\left(
 \frac{\cos\alpha}{\sin\beta}\sum_{i=1,2}[1\pm\frac{\sin2\theta_t}{2m_t}(A_t+\mu\tan\alpha)]\frac{m_t^2}{\tilde m_{ti}^2}F_{0}[\frac{4\tilde m_{ti}^2}{m_h^2}]\right. \nonumber
\\
&&\left.\hspace{1.7cm}-
 \frac{\sin\alpha}{\cos\beta}\sum_{i=1,2}[1\pm\frac{\sin2\theta_b}{2m_b}(A_b+\mu\cot\alpha)]\frac{m_b^2}{\tilde m_{bi}^2}F_{0}[\frac{4\tilde m_{bi}^2}{m_h^2}]
\right. \nonumber\\
&&\left. -
\sin(\alpha+\beta)\sum_{i=1,2}(\frac 12\{\begin{array}{c}  \cos^2\theta_t\\  \sin^2\theta_t\end{array}\}
\mp\frac 23\sin^2\theta_W\cos2\theta_t)\frac{M_Z^2}{\tilde m_{ti}^2}F_{0}[\frac{4\tilde m_{ti}^2}{m_h^2}]\right. \nonumber\\
&&\left.  +
\sin(\alpha+\beta)\sum_{i=1,2}(\frac 12\{\begin{array}{c}  \cos^2\theta_b\\  \sin^2\theta_b\end{array}\}
\mp\frac 13\sin^2\theta_W\cos2\theta_b)\frac{M_Z^2}{\tilde m_{bi}^2}F_{0}[\frac{4\tilde m_{bi}^2}{m_h^2}]\right)\!\!,
 \label{eq:squark_light}\\ 
\!\!\!\!\!\Delta{\cal M}_{H}&=&\frac{\alpha_s}{4\pi}  \frac{m_H^2}{2\sqrt{2}v}\left(
 \frac{\sin\alpha}{\sin\beta}\sum_{i=1,2}[1\pm\frac{\sin2\theta_t}{2m_t}(A_t-\mu\cot\alpha)]\frac{m_t^2}{\tilde m_{ti}^2}F_{0}[\frac{4\tilde m_{ti}^2}{m_H^2}]\right. \label{squark}\nonumber
\\
&&\left.\hspace{1.7cm}+
 \frac{\cos\alpha}{\cos\beta}\sum_{i=1,2}[1\pm\frac{\sin2\theta_b}{2m_b}(A_b-\mu\tan\alpha)]\frac{m_b^2}{\tilde m_{bi}^2}F_{0}[\frac{4\tilde m_{bi}^2}{m_H^2}]
\right. \nonumber
\end{eqnarray}
\begin{eqnarray}
&&\left. +
\cos(\alpha+\beta)\sum_{i=1,2}(\frac 12\{\begin{array}{c}  \cos^2\theta_t\\  \sin^2\theta_t\end{array}\}
\mp\frac 23\sin^2\theta_W\cos2\theta_t)\frac{M_Z^2}{\tilde m_{ti}^2}F_{0}[\frac{4\tilde m_{ti}^2}{m_H^2}]\right. \nonumber\\
&&\left.  -
\cos(\alpha+\beta)\sum_{i=1,2}(\frac 12\{\begin{array}{c}  \cos^2\theta_b\\  \sin^2\theta_b\end{array}\}
\mp\frac 13\sin^2\theta_W\cos2\theta_b)\frac{M_Z^2}{\tilde m_{bi}^2}F_{0}[\frac{4\tilde m_{bi}^2}{m_H^2}]\right)\!\!,
\label{eq:squark_heavy} 
\end{eqnarray}
\begin{eqnarray} 
 \label{eq:squark_pseudo}
\Delta{\cal M}_{A}&=&0 ,
  \end{eqnarray}
 \label{eq:squark}
 \end{subequations}
  where the squark mixing parameters and the mixing angles are 
  \beq
  X_b=A_b-\mu\tan\beta, \ \  X_t=A_t-\mu\cot\beta, \ \ \sin2\theta_q=\frac{2m_qX_q}{\tilde m_{q1}^2-\tilde m_{q2}^2}
  \eeq 
  and the upper sign corresponds to squark$_1$ and lower sign to squark$_2$.
  
  Note that due to the appearance of the quark mass squared versus $\tan\beta$, the main contribution comes from the t-squarks and not from the b-squarks. 
   The triangle  functions entering into Eqs.~(\ref{eq:quark},\ref{eq:squark}), $F_{1/2}^{h,H,A}$  and $F_0$, are given by Ref.~\refcite{okun}
 \begin{eqnarray}
              F_{1/2}^{h,H}[x]&=&-2x(1+(1-x)f(x)),\nonumber\\
               F_{1/2}^{A}[x]&=&-2x f(x),\nonumber\\
 F_{0}[x]&=&x(1-x f(x)), \label{eq:F}\\
 f(x)&=&\theta(x - 1) [ArcTan\frac{1}{\sqrt{x-1}}]^2 +  \theta(1 -x)  [\frac i2 Log(\frac{1 + \sqrt{1 - x}}{1 - 
              \sqrt{1 -x}}) + \frac \pi 2]^2 .\nonumber
              \end{eqnarray}
              
\begin{figure}[htb]\vspace{0.3cm}
\begin{center}
\includegraphics[width=6cm,keepaspectratio=true]{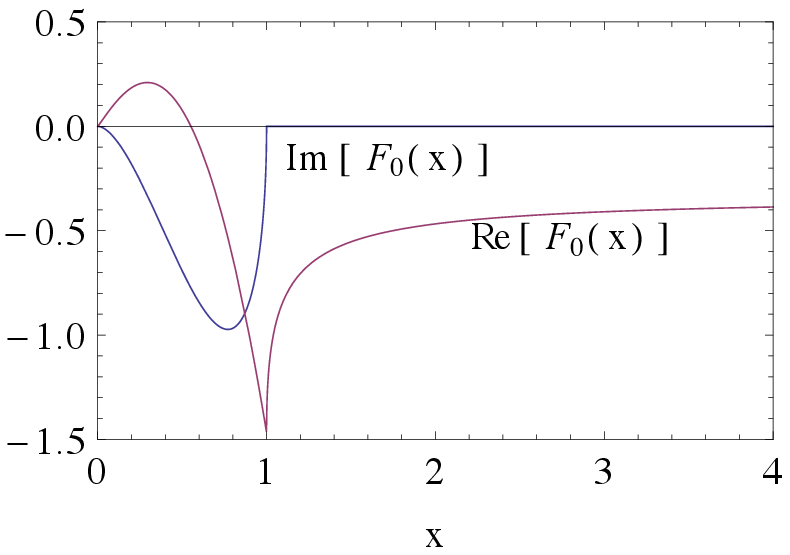}
\includegraphics[width=6cm,keepaspectratio=true]{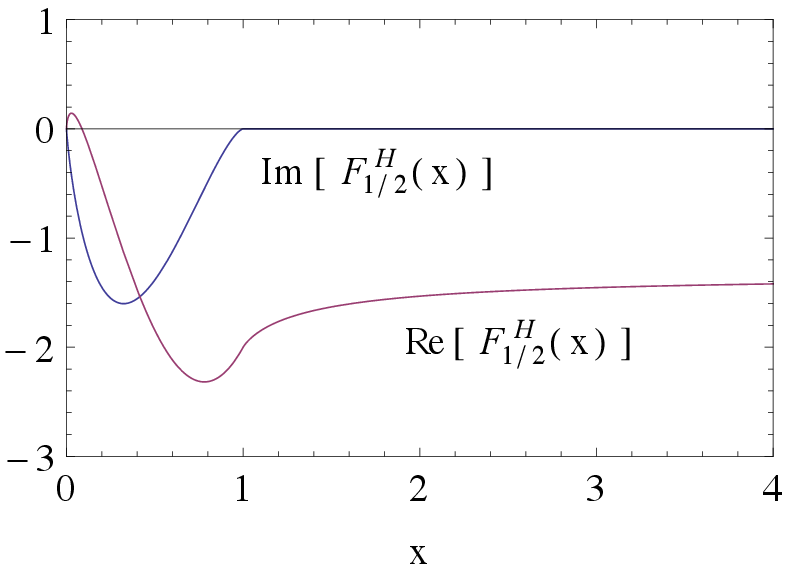}
 \caption{The dependence of the functions $F_0(x)$ and $F^H_{1/2}(x)$ on $x$.} 
\label{fig:f0_f12}
 \end{center}
 \end{figure}

The functions     $F_{1/2}^{h,H,A}$  and $F_0$ are complex functions of a single argument and get the imaginary part at the
threshold when $m_{q_i}=m_{Higgs}/2$ (see Fig.~\ref{fig:f0_f12}). 
At the threshold the modulus of $F_0$ is maximal and saturates the squark contribution. 
Thus, the desired enhancement of the cross-section is achieved at the threshold  when the mass of the t-squark and the heavy Higgs boson are correlated and differ by a factor of 1/2.  

In all the formulas the values of quark masses and $\alpha_s$ should be taken at the $m_{Higgs}$ scale.
In what follows all the needed low-energy running parameters are calculated with the help of SOFTSUSY 3.1.6\cite{Allanach:2001kg}  code which does not only 
perform the  RG evolution but also incorporates the important threshold effects, in particular, to the b-quark mass\cite{Carena:1999py}, which is essential for our analysis.

\section{Universal soft SUSY breaking framework} 
\label{sec:universal}

We start with the simplest mSUGRA-inspired scenario. Then in the MSSM with universal boundary conditions one has 4 parameters: $m_0,m_{1/2},A_0$ and $\tan\beta$. We take the sign of $\mu$ to be positive because of the SUSY contribution to $g-2$ of muon\cite{g-2}.  In what follows, we fix $\tan\beta$ to be large of the order of $30\div 50$.  This choice is motivated, on the one hand, by enhancement of the Higgs production cross-section and, on the other hand, by relic abundance of the DM
in the Universe interpreted as a SUSY WIMP\cite{Beskidt:2010va}. 
We present our results in the $m_0,m_{1/2}$ plane varying the values of $A_0$ and $\tan\beta$. As it will be clear later, the parameter $A_0$ has to be large, and it plays an essential role in the squark contribution. 

We have performed the calculations of the cross-sections according to Eqs.~(\ref{eq:sigma}-\ref{eq:squark}) with the MSTW2008-LO gluon distribution function\cite{PDFs1,PDFs2,PDFs3} taken at $Q\sim m_A \sim m_H$. 
It is known that the leading order result can be substantially modified by the inclusion of high order (S)QCD corrections\cite{Spira:1993bb}. The net effect of the NLO\footnote{In fact, NNLO\cite{nnlo1,nnlo2} and even N$^3$LO\cite{n3lo1,n3lo2,n3lo3,n3lo4} QCD corrections to quark loops are available today.} diagrams is usually summarized in the form of the so-called K-factors
\begin{equation}
	K = \frac{\sigma_{\mbox{NLO}}}{\sigma_{\mbox{LO}}}.
\label{eq:K-factor}
\end{equation}
For small values of $\tan\beta$ the K-factor can enhance the cross-section by 100 \%. However, it turns out that in the case of high $\tan\beta$ the K-factors for heavy Higgs bosons are much smaller ($\mathrm{K}=1.1-1.2$) and comparable with the overall theoretical uncertainty of the NLO result\cite{Muhlleitner:2006wx}. 
Since we are looking for enhancement by a factor of several units we ignore these subtleties here 
albeit of their importance in precision analysis.

As it was mentioned above, to calculate the spectrum of superpartners and the other low scale parameters from the high energy ones, we use
the RG running implemented in SOFTSUSY 3.1.6 code\cite{Allanach:2001kg}. As the benchmark points,  we choose
three points in the $m_0,m_{1/2}$ plane to be  $(m_0,m_{1/2})=(900,300)$~GeV,  $(m_0,m_{1/2})=(1100,300)$~GeV, and $(m_0,m_{1/2})=(1700,200)$~GeV, respectively, and allow $A_0$ to be positive and negative.  This choice is dictated, on the one hand, by the requirement of smallness of $\tilde{m}_{t1}$ which gives the main contribution to the cross-section and, on the other hand, by restrictions  on the parameter space coming from the other physical constraints\cite{Kaz}.

\begin{figure}[H]
\begin{center}
\includegraphics[width=6cm]{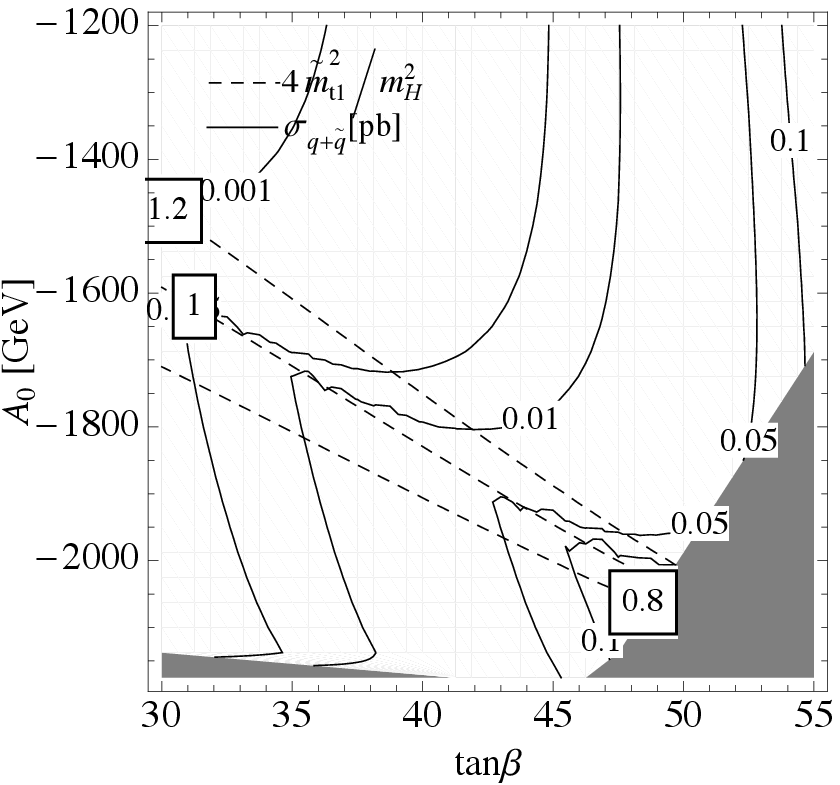}
\hspace{0.2cm}
\includegraphics[width=6cm]{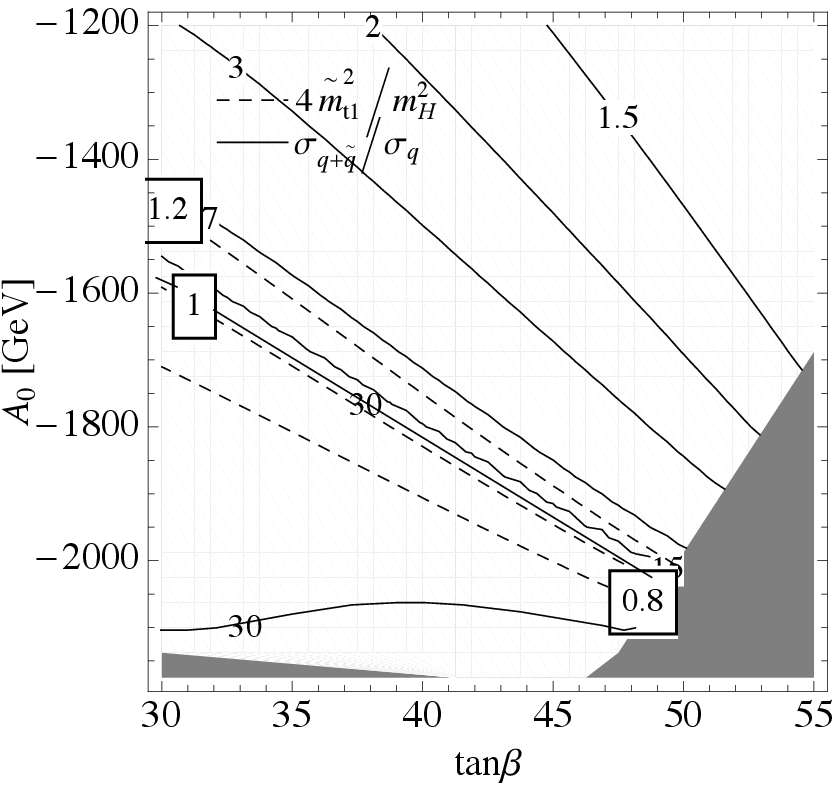}
\includegraphics[width=6cm]{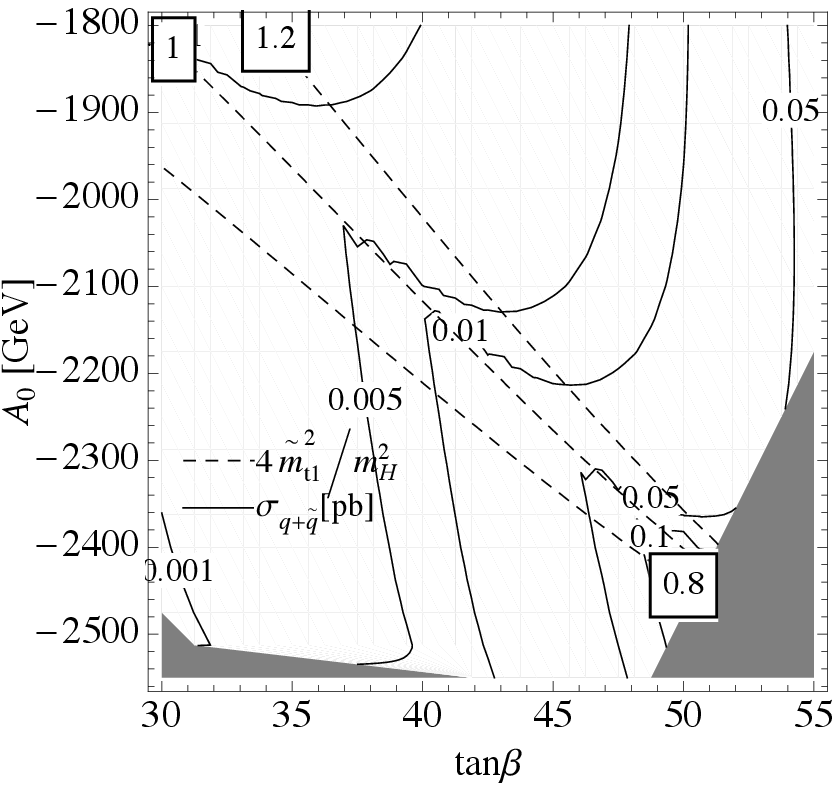}
\hspace{0.2cm}
\includegraphics[width=6cm]{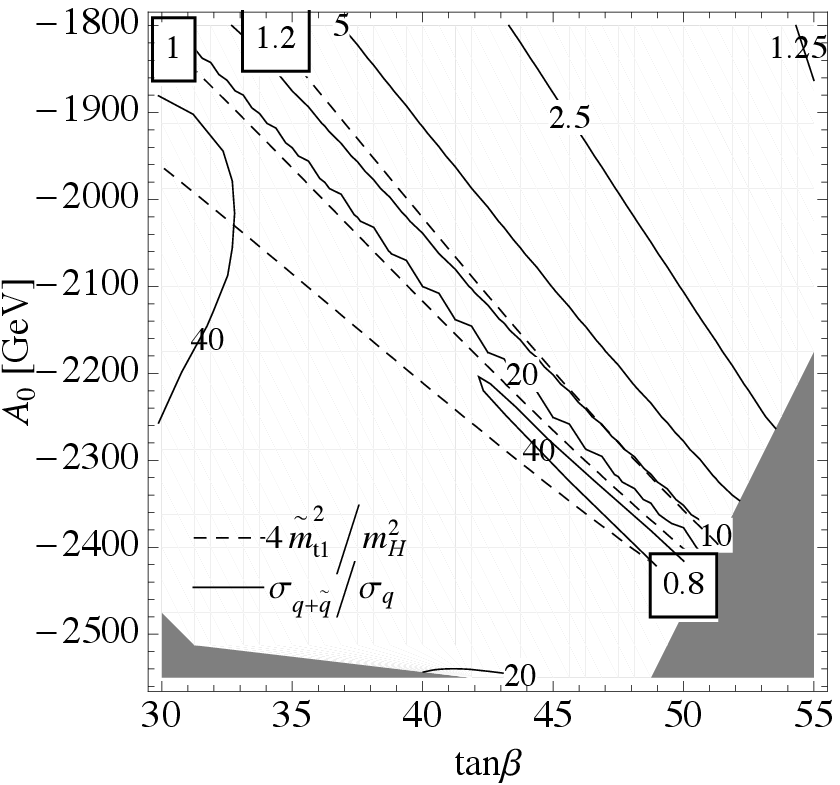}
\includegraphics[width=6cm]{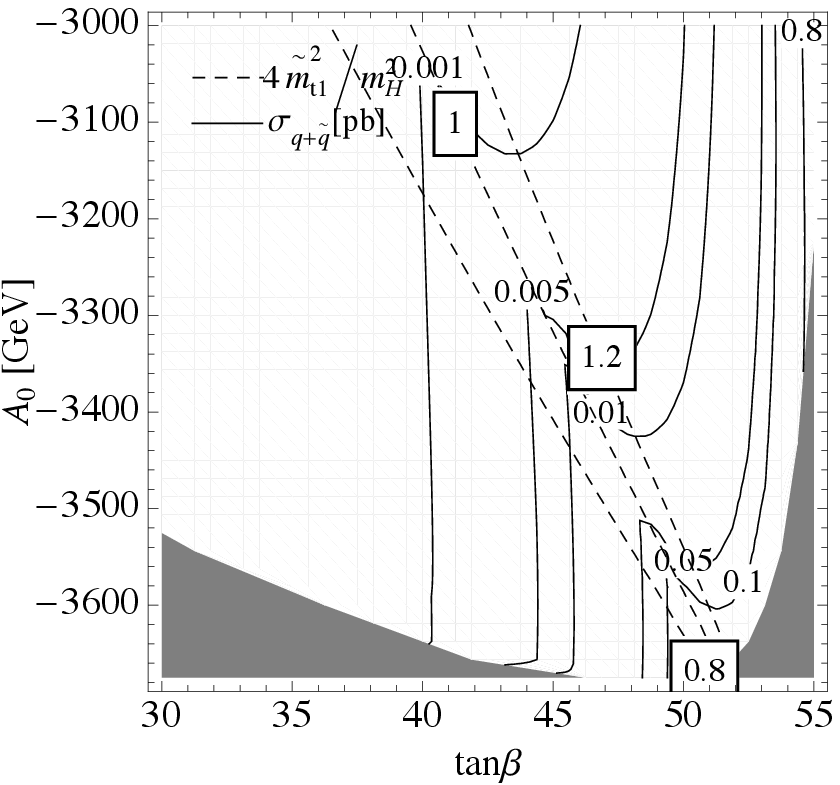}
\hspace{0.2cm}
\includegraphics[width=6cm]{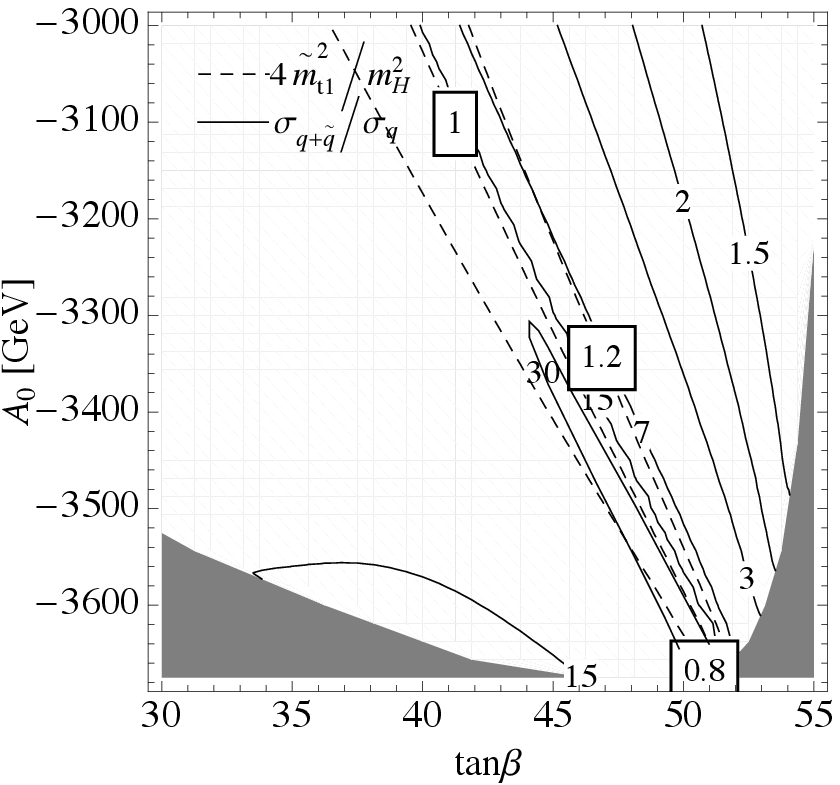}
 \caption{
  The cross-section ($\sqrt s = 14$~TeV) of the heavy Higgs production (left) and the ratio of squark+quark  to quark loop  contribution (right) as functions $A_0$ and $\tan\beta$  for the points $(m_0,m_{1/2})=(900,300)$~GeV,  $(m_0,m_{1/2})=(1100,300)$~GeV,and  $(m_0,m_{1/2})=(1700,200)$~GeV, respectively.  
The dashed lines correspond to the resonance values of $4\tilde{m}^2_{t1}/m^2_H$. At the threshold $4\tilde{m}^2_{t1}/m^2_H=1$ and the enhancement is maximal.  The gray regions are prohibited by the LSP constraint or due to the existence of a tachyon in the parameter space.
\label{fig:contour} }
 \end{center}
 \end{figure}

The total cross-section  for the heavy Higgs boson production as well as the ratio of the quark+squark cross-section to the quark one for three different benchmark points are shown in Fig.~\ref{fig:contour}.
The most significant contribution to the production cross-section, $\sigma_{q+\tilde q}$, comes from the loop with the lightest squark $\tilde t_1$ which gives almost 99\%  of the total value. The contribution of $\tilde t_2$ is suppressed by its heavy mass $m_t^2/\tilde{m}^2_{t2}$ and those of the b-squarks by the ratio $m_b^2/\tilde{m}^2_b$. 
The desired enhancement due to the squark contribution is achieved via the terms in Eq.~(\ref{squark}) proportional to $\sin 2\theta_t$.  
The big enhancement can be obtained only for large and negative values of $A_0$. 
One can understand qualitatively the result by noticing that the soft triple coupling $A_t$ which starts at $A_0$ at high scale tends to the IR fixed point at low energy\cite{CK}  which is always negative. As a result, the absolute value of $A_t$ is minimal for positive $A_0$ and maximal for negative $A_0$. At the same time, the stop mixing is proportional to $A_t$ and the bigger the mixing the smaller is the top squark mass $\mms{t}{1}$  
and, hence, the bigger is the cross-section. So negative values of $A_0$ are favourable.
Taking $A_0$ to be big and negative it is possible to get the total cross-section to be 
of the order of 0.1 pb with the enhancement factor due to squarks of the order of several units.

One can clearly see in Fig.~\ref{fig:contour} that the highest values of the ratio $R_H\equiv \sigma_{q+\tilde{q}}/\sigma_q$ and the total cross-section $\sigma_{q+\tilde{q}}$ are achieved along the  straight lines  which correspond to the resonance with  $4\tilde{m}_{t1}^2/m_H^2=1$. 
 This is due to the properties of the functions $F_0$ and $F^H_{1/2}$ mentioned above and the fact that the squark and
 quark amplitudes interfere constructively at the threshold for $A_t <0$.
 
As a result, one gets considerable enhancement of the cross-section with the leading role played by the lightest stop in the loop.  The total cross-section reaches a fraction of pb that opens the  possibility of earlier Higgs boson detection.
The weak point of our analysis is the necessity  of large negative values of $A_0$ which seems to contradict the fits\cite{Beskidt:2010va}
to the $b\to s\gamma$
decay rate for large $\tan\beta$. It turns out that for the considered regions 
$\mathrm{BR}(B\to X_s \gamma)\simeq 10^{-5}$, which is an order of magnitude lower than the experimental value\cite{Barberio:2008fa}   
$(3.55\pm0.24\pm0.09)\times 10^{-4}$.
A careful investigation of the problem shows that for negative $A_0$ the chargino-stop contribution\cite{Barbieri:1993av,Degrassi:2000qf}  to the
Wilson coefficient $C_7$\footnote{that corresponds to the effective operator $O_7 = e^2/(16\pi^2) m_b (\bar s_L \sigma^{\mu\nu} b_R) F_{\mu\nu}$.}, which influences the $b\to s\gamma$ rate at the leading order, tends to cancel the contributions due to charged Higgs and W-boson.  
In the considered scenarios the correction $C^\chi_7$ has the same order of magnitude as the sum $C^W_7 + C^H_7$ from charged Higgs and W-boson.
Since $\mathrm{BR}(B\to X_s \gamma)_{\mathrm{LO}}\propto |C_7|^2$, one can immediately deduce that the corresponding branching ratio is lower than that of the SM. 
Moreover, it turns out that the constraint due to the DM relic density is also hard to 
fulfill in the considered regions, and the non-observation of $B_s\to\mu^+\mu^-$ (see Refs.~\refcite{Bmumuexp,Bmumuexp2}) forbids the most promising part of the plane with $\tan\beta\gtrsim 45$. 

To overcome the above-mentioned difficulties with  $b\to s\gamma$, one can consider positive $A_0$ in which case all the corrections to $C_7$ 
have the same sign and there is a good chance to have a proper value of the branching ratio.
However, in the universal SUSY breaking scenarios with positive $A_0$ it is impossible to have simultaneously large SUSY enhancement and a high heavy Higgs production cross-section.
For example, choosing the low value of $m_{1/2}\simeq 200$~GeV, moderate $m_0\simeq 500-600$~GeV, and $\tan\beta\simeq 25-30$ 
one can reduce the lightest stop mass below 200~GeV with the help of large $A_0\simeq 2000$~GeV. 
However,  for this set of parameters the Higgs mass $m_H$ is too large, i.e., $m_H \gg \ms{t}{1}$, and, consequently,  the total cross-section is very small.  

The other possibility  would be significant enhancement of the chargino-stop loop so that $|C^\chi_7|$ is an order of magnitude larger than $|C^W_7+C^H_7|$.  This effect strongly depends on the value of the $\mu$-parameter which influences the masses and mixing of charginos. In the mSUGRA parameter regions considered 
here  we have $\mu\sim 1$~TeV and  $C^\chi_7$ is suppressed. 

Clearly, to save the situation and to get a reasonable phenomenological impact from the squark contribution 
we are forced to release the universality constraint for the Higgs masses and consider the NUHM model\cite{Ellis:2002wv}.

\section{Non-universal soft supersymmetry breaking}
\label{sec:non_universal}

The non-universality in the Higgs sector parameterized by the pole mass of the CP-odd heavy boson $m_A$ and
the running $\mu$-parameter at the SUSY scale (NUHM scenario) provides us with the possibility to overcome 
the above-mentioned difficulties of the universal scenarios for both negative and positive $A_0$.
For the large $\tan\beta$ scenarios $m_H \simeq m_A$ and we have enough freedom to obtain 
significant enhancement in the $\sigma_{gg\to H}$ cross-section by adjusting $m_A$. Moreover, since we also 
can adjust the $\mu$-parameter, it is possible to fulfill the $b\to s\gamma$ constraint by the increase of the chargino 
contribution mentioned in the previous section.

In the first place, we consider the case with negative $A_0$ and try to find a region that satisfies all the above-mentioned 
experimental constraints.

In order to enhance the chargino contribution to the $b\to s\gamma$ decay rate, we need to decrease the value of the $\mu$ parameter. 
This, in turn, lowers the scale of stop masses.  As a consequence,  the lightest stop can become an LSP or even a tachyon if we consider very large values of $|A_0|$ which were chosen in the previous section. This kind of reasoning justifies our choice of $A_0$ given below.

\begin{figure}[t]
	\begin{center}
		\includegraphics[width=4cm]{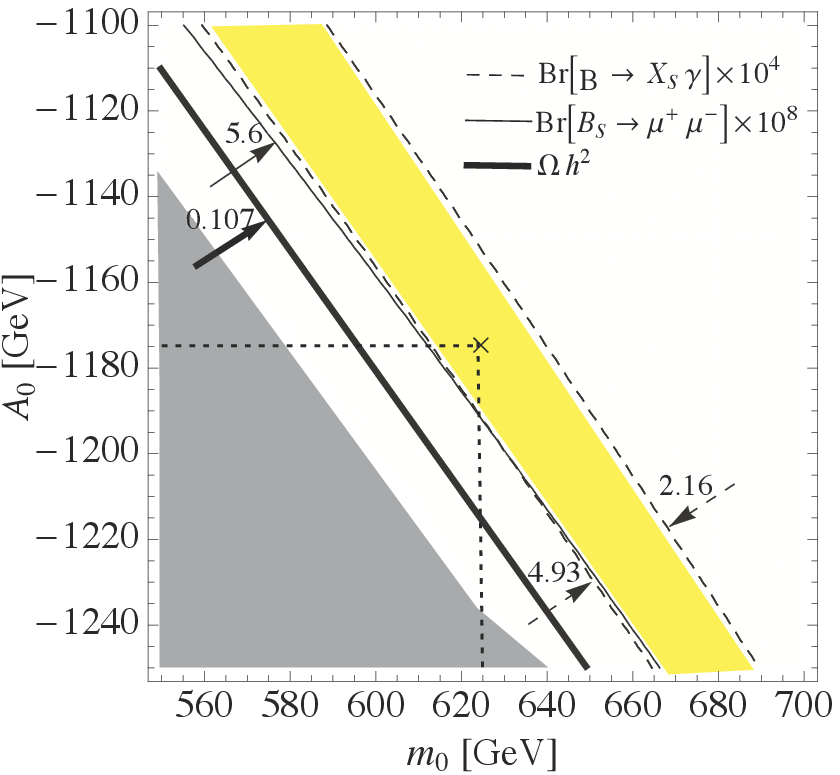}
		\includegraphics[width=4cm]{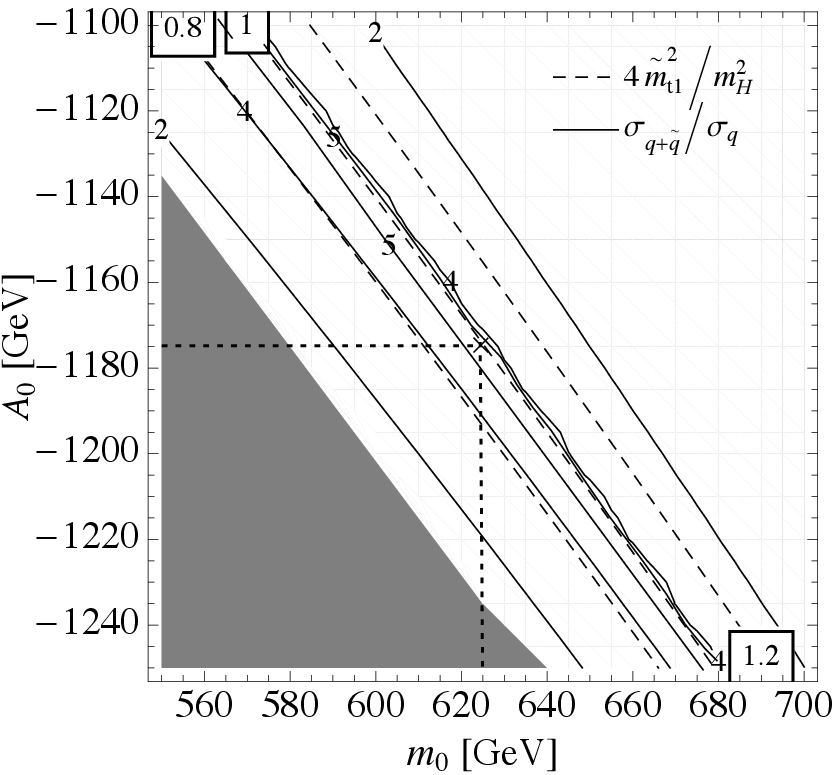}
		\includegraphics[width=4cm]{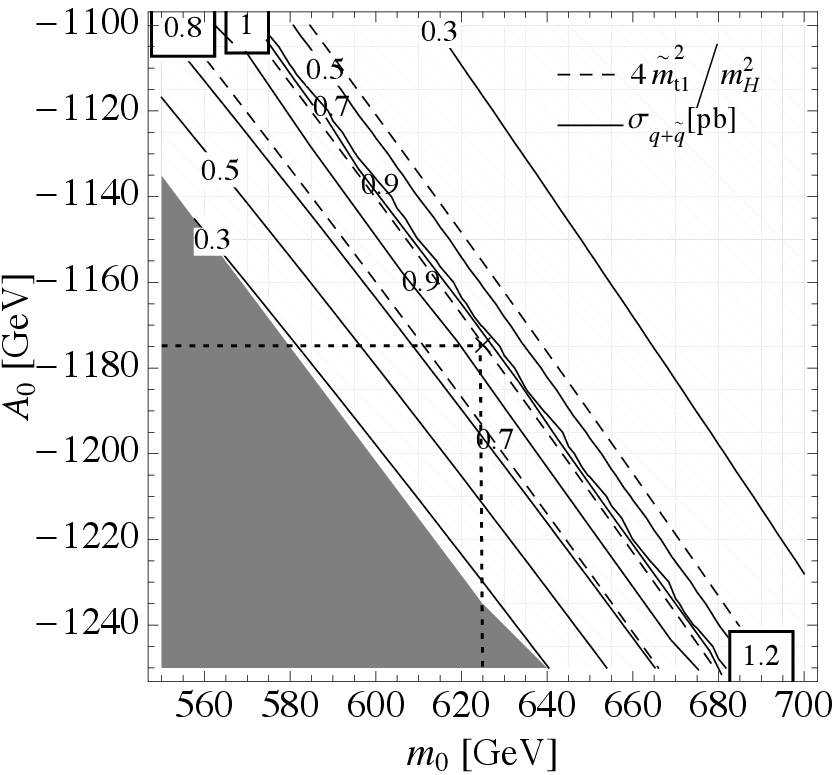}
	\end{center}
	\begin{center}
		\includegraphics[width=4cm]{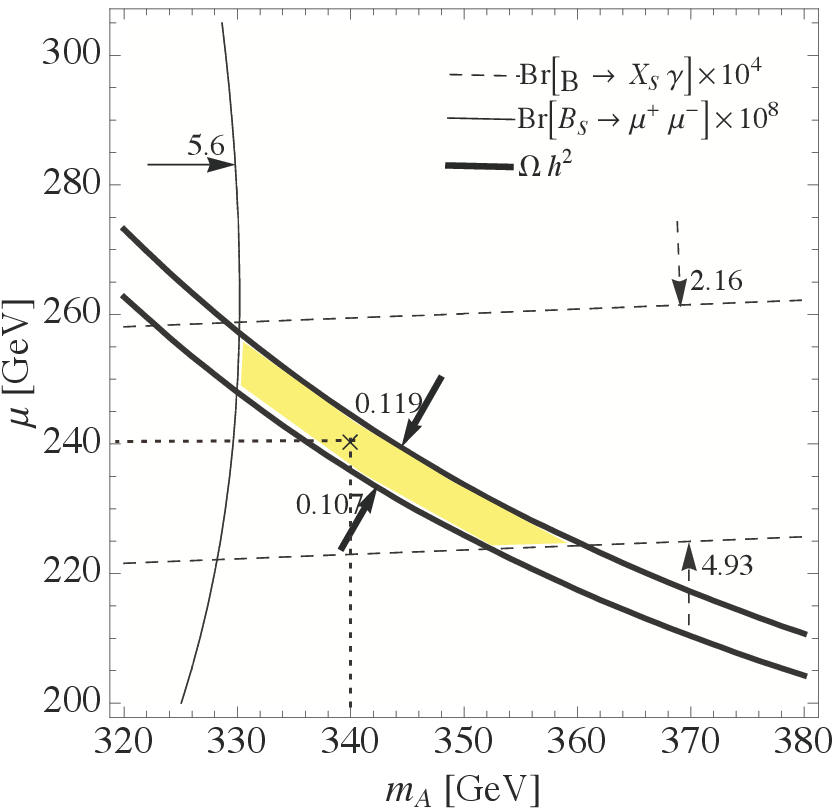}
		\includegraphics[width=4cm]{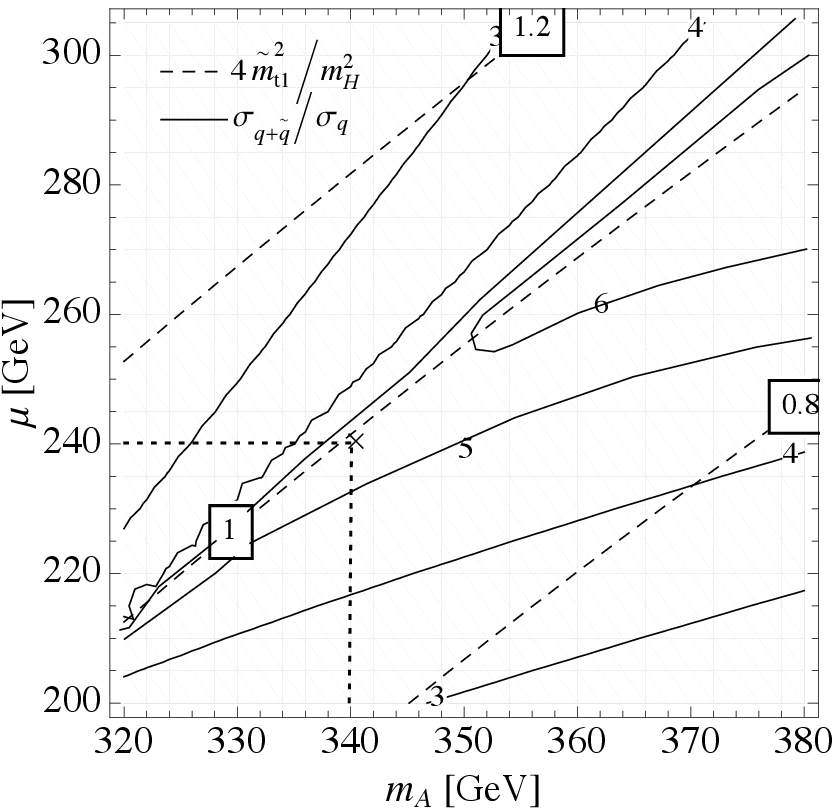}
		\includegraphics[width=4cm]{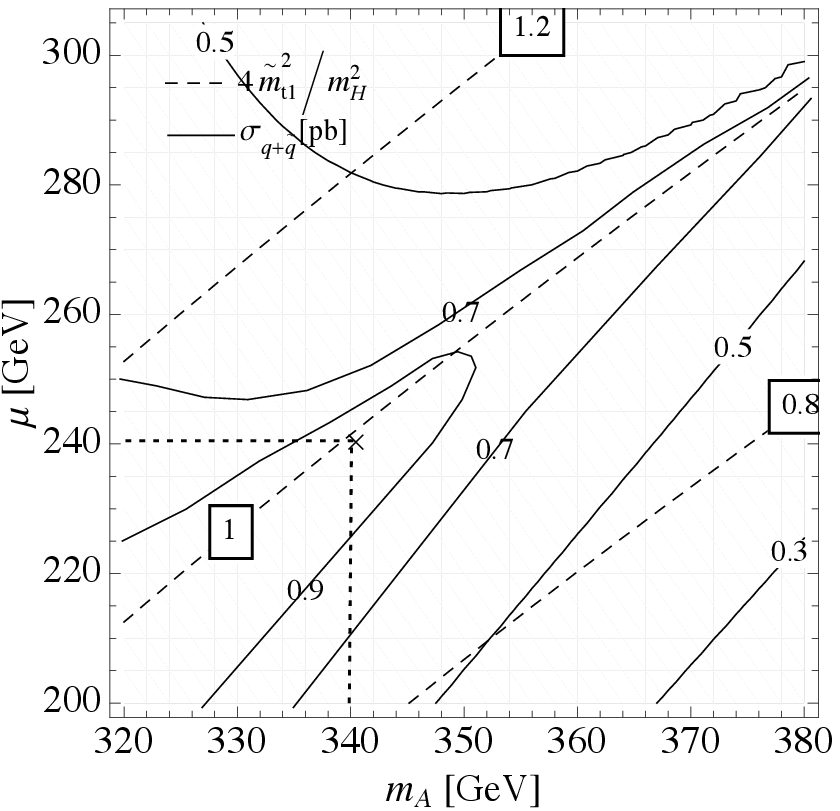}
	\end{center}
	\begin{center}
		\includegraphics[width=4cm]{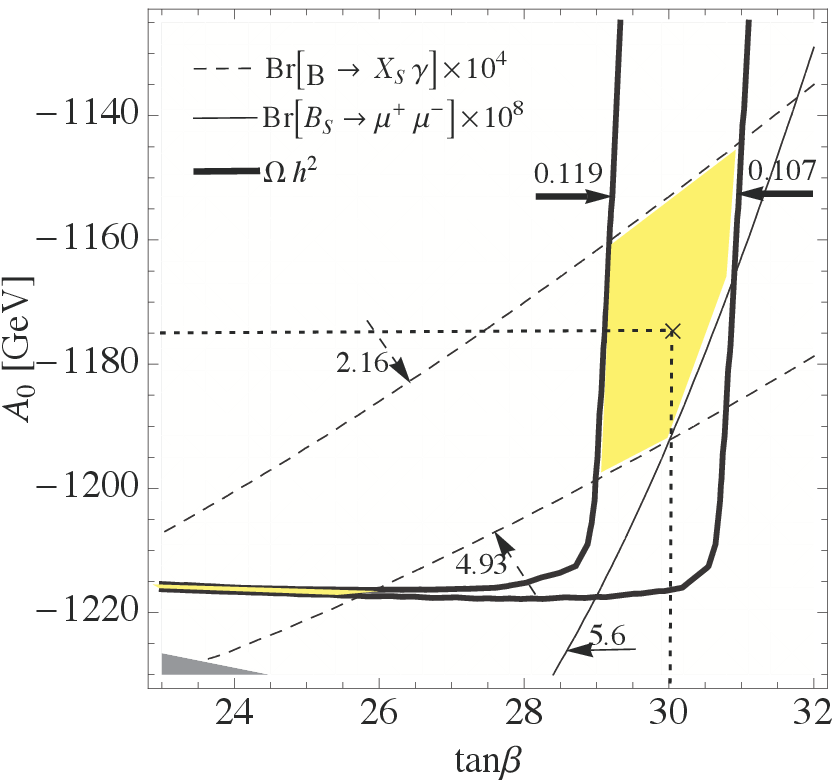}
		\includegraphics[width=4cm]{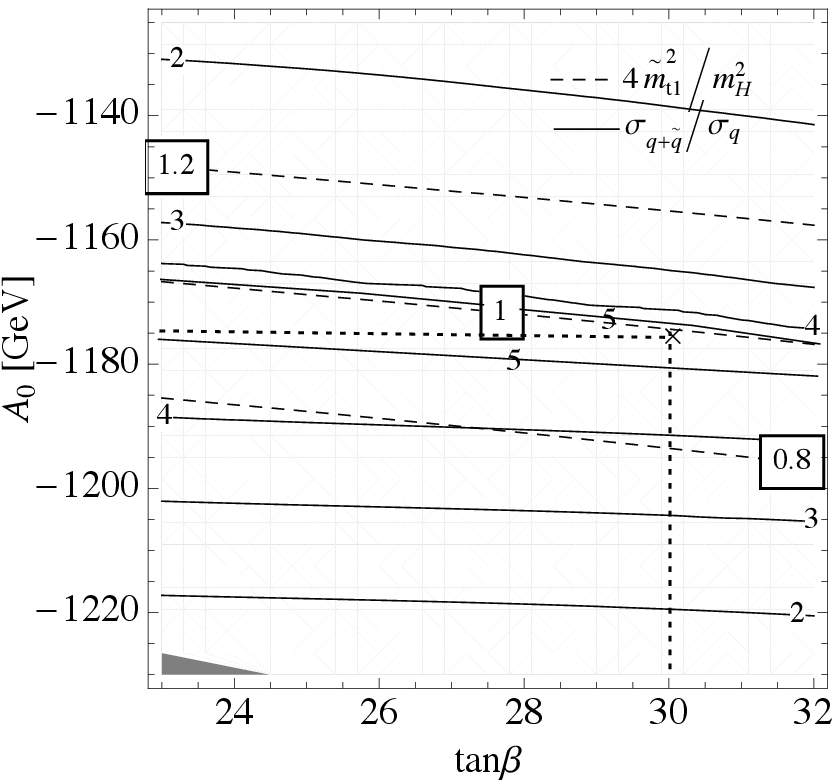}
		\includegraphics[width=4cm]{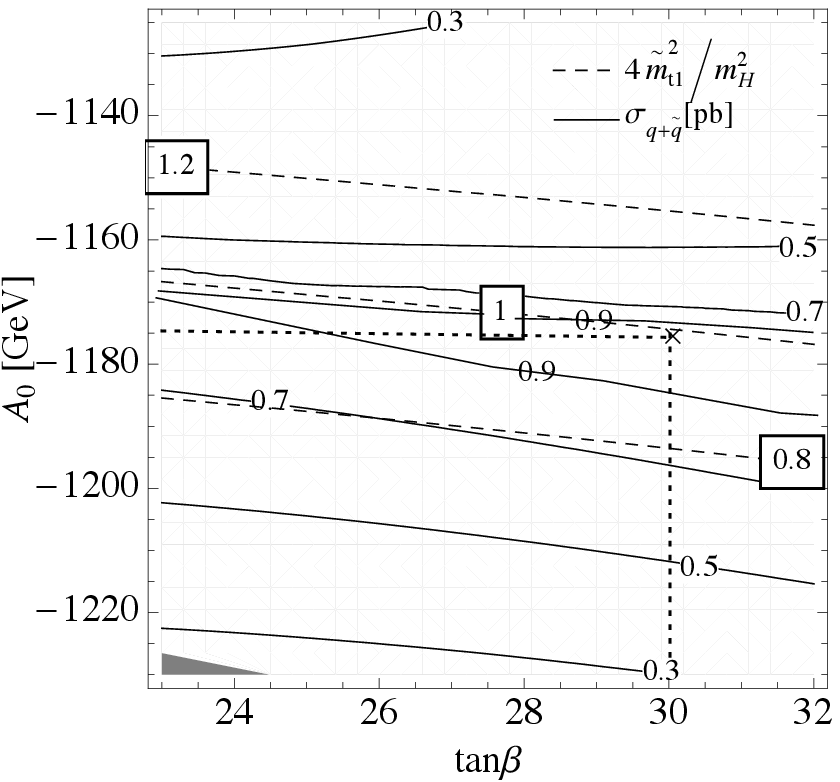}
	\end{center}
	\caption{The allowed regions in  the $m_0-A_0$ plane for $\tan\beta=30$, $m_{1/2}=250$ GeV, $m_A=340$ GeV, $\mu=240$ GeV (top), $m_A - \mu$ plane for $\tan\beta=30$, $m_0=625$ GeV, $m_{1/2}=250$ GeV, $A_0=-1175$ GeV (middle) and
$\tan\beta - A_0$ plane for $m_0=625$ GeV, $m_{1/2}=250$ GeV, $m_A=340$ GeV $\mu=240$ GeV (bottom). On the left panel, all of the regions between the arrows are allowed by the $B_S\rightarrow\mu^+\mu^-$, $B\rightarrow X_S\gamma$ and $\Omega h^2$ constraints. The intersection of these regions is marked by  yellow. 
The ratio of the cross-sections $R_{H} = \sigma_{q+\tilde q}/\sigma_{q}$ and the total cross-section $ \sigma_{q+\tilde q}$ at $\sqrt s = 14$~TeV are shown on the middle and the right panels, respectively. The numbers  $0.8, 1.0$ and $1.2$ on the middle and right panels correspond to the values of the ratio $4\tilde{m}_{t1}^2/\tilde{m}_{H}^2$.  The benchmark point is marked by a cross.}
	\label{fig:non_uni_m0A0}
\end{figure}

As a benchmark point we have used the following set of NUHM parameters $m_{1/2} = 250$~GeV, $m_{0} = 625$~GeV, $\mu = 240$~GeV, $m_A = 340$~GeV, 
$A_0 = - 1175$~GeV, $\tan\beta = 30$. This point lies in the region bounded by the experimental constraints mentioned above. 
Obviously, it is very hard to visualize the allowed region in the space of six free parameters. 
In what follows, we present  in Fig.~\ref{fig:non_uni_m0A0} the two-dimensional sections of the region in 
$m_0 - A_0$, $m_A-\mu$  and  $\tan\beta-A_0$  planes, respectively.  
One can see how the allowed regions due to various constraints intersect with each other. For the calculation of
the flavour observables and the relic density we use the SuperIso (Relic) code\cite{superiso1,superiso2,superiso_relic}, and the bounds 
for $b\to s\gamma$\footnote{The allowed interval for $b\to s\gamma$ includes also theoretical uncertainties, 
see, e.g., Ref.~\refcite{superiso1}.},
$B_s\to\mu^+\mu^-$, and $\Omega h^2$ correspond to 95 \% CL. A point marked by the cross corresponds to the chosen benchmark scenario.

This choice is somewhat random within the allowed region. Looking at the plot in Fig.~\ref{fig:non_uni_m0A0}
one can see where the allowed region moves when varying one or more parameters. 
For example, looking at the $A_0-\tan\beta$ plane one can deduce that with a slight increase of $\tan\beta$ both the $b\to s\gamma$ 
and the $B_s\to\mu^+\mu^+$ rates go up and the allowed strips in the $m_0-A_0$ and $m_A - \mu$ planes 
effectively move towards the lower values in the corresponding figures.  
However, the correlations between the degrees of freedom are strong enough, so that it is hard 
to get the entire picture. In what follows we try, at least qualitatively, 
to explain the key features of the emerged picture.

Since the stop mass scale depends crucially on the $m_{1/2}$ parameter, we restrict ourselves 
to the  value  $m_{1/2} = 250$~GeV. All the other parameters are allowed to vary. 
It turns out that the constraints due to the muon anomalous magnetic moment and 
electroweak precision data\footnote{We use FeynHiggs 2.7.4\cite{Hahn:2010te}  to evaluate $\Delta \rho$ 
that parametrizes the leading universal corrections to the electroweak precision observables.}   
are satisfied in the whole region studied 
($1 \lesssim a_\mu \times 10^9  \lesssim 2.5$, $\Delta \rho \lesssim 5\cdot 10^{-4}$), so we do not draw the corresponding bounds.

In the same figures, the SUSY enhancement of the Higgs production via
the gluon fusion is demonstrated with the help of the ratio $R_{H} = \sigma_{q+\tilde q}/\sigma_{q}$.
Clearly, due to the fact that the quark contribution for our case is not very small, 
the enhancement is not very big in comparison 
with the results presented in the previous section, e.g., $R_H \sim 5$ for the benchmark point. Again, the value of $R_H$
correlates with $\xst1 \equiv 4\mms{t}{1}/m_H^2$. At the lightest stop production threshold it is maximal and $R_H$ decreases
more rapidly when $\xst1 > 1$. In spite of the moderate enhancement the total cross-section is of the order of pb at the stop production threshold.

At the top of Fig.~\ref{fig:non_uni_m0A0} the plane $m_0-A_0$ is shown for fixed $\tan\beta = 30$, $m_{1/2} = 250$~GeV, $m_A=340$~GeV, and $\mu=250$~GeV.
One can see that  the parameters $A_0$ and $m_0$ are correlated within the allowed band. 
This correlation corresponds to
a constant value of the lightest stop mass lying in the range $150-200$~GeV and can be easily explained by the fact that an increase in 
the stop mass  with $m_0$ can be compensated via a see-saw like mechanism by an increase in the off-diagonal term in the stop mass matrix driven by the absolute value of $A_t$. Clearly, both the $B\to \mu^+\mu^-$ and $B\to X_s \gamma$ rates go down with $\ms{t}{1}$.

In the middle of Fig.~\ref{fig:non_uni_m0A0}, we show  how the allowed bands due to various constraints intersect in the $m_A - \mu$ plane. 
One can notice the dependence of the $b\to s \gamma$ rate on $\mu$ which somehow supports 
our hypothesis about the dominance of the chargino contribution to $C_7$ Wilson coefficient for small $\mu$. 
With the increase of $m_A$ the charged Higgs mass increases correspondingly. As a consequence, the sum $C^H_7 + C^\chi_7$ becomes bigger, 
thus, slightly increasing the branching fraction.

The correct amount of the Dark Matter can be achieved if LSP annihilates via the virtual CP-odd Higgs boson in the s-wave. For
this to happen, the neutralino mass $m_{\chi^0}$ should be adjusted to half $m_A$. In our case, for fixed $m_{1/2}=250$~GeV the neutralino is mostly
bino with $m_{\chi^0}$ around 100~GeV. Moreover, if $\mu$ is comparable with $m_{1/2}$ the fraction of higgsino component in $\chi^0$ becomes larger
and also increase the cross-section which is proportional to the mixing between the gaugino and higgsino components for the s-wave annihilation. 
These two facts explain, at least qualitatively, the behavior of the curves with constant value of the DM relic density. 
For low $\mu\sim 200$~GeV it is sufficient to have $m_A \simeq 400$ GeV to obtain the correct value of $\Omega h^2$. 
However, when due to the increase of $\mu$ the mixing between the  gaugino and  higgsino components becomes small,  
one needs to lower $m_A$ to be closer to the $A_0$-resonance to enhance the annihilation cross-section. 
For the considered value of $\tan\beta=30$ the upper bound from $B_s \to \mu^+\mu^-$ excludes $m_A\lesssim 330$~GeV. All the constraints are 
satisfied in the small region near our benchmark point.

Finally, at the bottom of  Fig.~\ref{fig:non_uni_m0A0} the $\tan\beta-A_0$ plane is shown. It is easy to notice that large $\tan\beta\gtrsim30$ are excluded by the 
$B_s \to \mu^+\mu^-$ constraint since the dominant SUSY contribution to this decay scales as $\tan^6\beta$.\cite{Bobeth:2001sq}  
In the allowed strip due to the $b\to s\gamma$ constraint the parameters $A_0$ and $\tan\beta$ are correlated since the enhancement due to $\tan\beta$ 
is compensated by the increase of $\ms{t}{1}$ due to $A_0$. The relic density constraint fixes $\tan\beta$ to be around 30. A tail of the $\Omega h^2$ 
region corresponds to the stop co-annihilation.

In summary, the key features of the allowed region are the following: $m_{1/2} \sim \mu \sim 250$ GeV
(which influence significantly the lightest stop mass, $b\to s\gamma$, and the mass and content of the lightest neutralino), 
$\tan\beta \sim 30$ (mostly due to the $\Omega h^2$ constraint), $m_0$ and $A_0$ should be correlated (due to the stop mass), 
and $m_A\gtrsim 300$ should not be very large (to have the Higgs production cross-section at the level of 1 pb).
For our benchmark point the heavy CP-even Higgs boson decays predominatly into the heavy down-type fermions, 
i.e., $b\bar b$ ($\sim 90\%$) , $\tau\bar \tau$ ($\sim 10 \%$). The latter signature has already been analyzed by
both the ATLAS\cite{Schumacher:2011jq} and CMS\cite{Chatrchyan:2011nx} collaborations and important bounds 
on $m_A$ and $\tan\beta$ were deduced. However, the scenarios with $m_A>300$~GeV and $\tan\beta<50$ are not excluded at the moment.

Before going to conclusions let us mention the situation with the case of $A_0>0$. 
It should be noted that contrary to the $A_0<0$ case,  positive $A_0$ leads to destructive interference between the squark and quark amplitudes at the stop threshold in the cross-section for heavy neutral Higgs production. 
The only possibility to enhance the cross-section is to be slightly below the threshold
$\ms{t}{1} \lesssim m_H/2$ when the corresponding squark amplitude develops a negative imaginary part.
If we choose $m_A$ to be around 350-400~GeV, the SUSY enhancement with $R_H \sim 10$ is possible for $\ms{t}{1} \simeq 110$ GeV.
However, due to the behaviour of RGE for $A_t$, the large initial values of $A_0>0$ lead to a relatively small positive $A_t$ at the SUSY scale. In 
order to obtain the light stop needed for large $R_H$ via the see-saw like mechanism,  the overall stop mass scale 
should not be very big. 
Unfortunately, this latter fact prevents us from finding a suitable region in the parameter space with $A_0>0$, since it turns out
that for a setup like this the lightest Higgs boson mass is around 100~GeV, which is excluded experimentally 
(we use HiggsBounds 2.0\cite{Bechtle:2011sb} package for confronting our predictions with the LEP bound). 
In contrast, for the $A_0<0$ scenario we have $m_{h_0} \simeq 118$ GeV.

\section{Discussion}

The search for the Higgs boson seems to be the main goal for the LHC today though the appearance of the new physics would be the major breakthrough. One can see that even if the "new physics" is represented by the enlargement of the Higgs sector, 
the cross-section of the Higgs production can be essentially enhanced due to the large value of $\tan\beta=v_2/v_1$. This enhancement might even lead to   preferable observation of a heavy Higgs boson rather than the light one. 

At the same time, if SUSY or some other heavy particles exist, the enhancement of the Higgs production can be pushed even further. This latter enhancement, however, is valid only for the restricted set of parameters subjected to two requirements: one of the intermediate particles (the lightest top squark $\tilde t_1$ in our case) has to be relatively light and has to be close to the resonance with the Higgs boson.

The allowed region in the parameter space found here seems to be very narrow mostly due to the relic density constraint. However, this impression is not true since in each plane shown in Fig.~\ref{fig:non_uni_m0A0} all the other parameters are fixed.
In the whole parameter space the allowed volume with $\sigma_{q+\tilde q} \lesssim 1$ pb and $R_H\simeq 3-5$ 
is obviously bigger. For example, the benchmark point parameters can be shifted to $\tan\beta = 25$ and $\mu = 210$~GeV
at the price of slightly lower values of $R_H \sim 3$ and $\sigma_{q+\tilde q} \sim 0.5$ pb.

\begin{figure}[htb]
	\begin{center}
		\includegraphics[width=12cm]{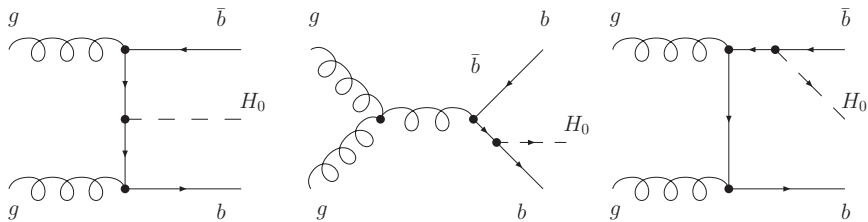} 
	\end{center}
	\caption{The LO diagrams for the associated heavy Higgs production with the two b-jets in the so-called four-flavor scheme (4FS), 
	where one does not consider b-quarks as partons in the proton.
	For large $\tan\beta$ this contribution to the total production cross-section is dominant.}
	\label{fig:bbH}
\end{figure}

Our main goal was to study the influence of squarks on the heavy Higgs boson production
and to find the regions of the MSSM parameter space, for which 
the cross-section via the gluon fusion process can be essentially increased. 
However, in the considered scenarios compatible with known experimental constraints 
it is still lower than the associated production accompanied by two b-quarks\cite{Djouadi:2005gj} (see diagrams shown in Fig.~\ref{fig:bbH}).

Indeed, with the help of CalcHep\cite{Pukhov:2004ca} package the total cross-section for $pp\to \bar b b H$ process
is estimated to be around 7 pb at $\sqrt s = 14$ TeV\footnote{Comparison is made with bbh@nnlo package\cite{Harlander:2003ai} and a reasonable agreement is found.} for our benchmark point ($\tan\beta = 30$, and $M_A = 340$ GeV). 
This is an order of magnitude larger than the gluon fusion cross-section evaluated above. Hence, it is very hard to ``see'' the gluon-fusion on top of the $b \bar b H$  process. 
It should be pointed out, however, that there are no virtual superpartners in the diagrams in Fig.~\ref{fig:bbH} 
so the same cross-section is expected  within any Two-Higgs Doublet Model (THDM) with large $\tan\beta$. 
As a conseqence,  a complimentary search is required to discriminate between different THDM possibilities. 

It is worth mentioning the other phenomenological implications of the chosen benchmark point with $A_0 < 0$. 
In the considered case the lightest top squark is almost degenerate with the top quark and its dominant decay 
channel is $\tilde t_1 \to \chi^+_1 b$ (we use SUSYHIT code\cite{Djouadi:2006bz} to calculate the branching fractions). 
This mode was not so extensively analyzed at the Tevatron and the current
bounds for the stop production at $\sqrt s = 1.96$ TeV are far above the theoretically predicted values\cite{Abazov:2009ps}.
However, at the LHC they can be produced abundantly.  For example, for our benchmark point 
the stop pair production cross-section at $\sqrt{s} = 14$~TeV that was obtained with the help of the Calchep package\cite{Pukhov:2004ca} 
is around 55 pb (in comparison with approximately 8 pb for $\sqrt{s} = 7$~TeV). 
The lightest chargino $\chi^+_1$ produced in the stop decay has the mass slightly below the neutralino-W-boson threshold 
($m_{\chi^+} \simeq 170~\mathrm{GeV} \lesssim m_{\chi^0} + m_W$), so it decays into the
lightest neutralino and a fermion-anti-fermion pair coming from the virtual $W$-boson.

\begin{figure}[htb]
	\begin{center}
		\includegraphics[width=7cm]{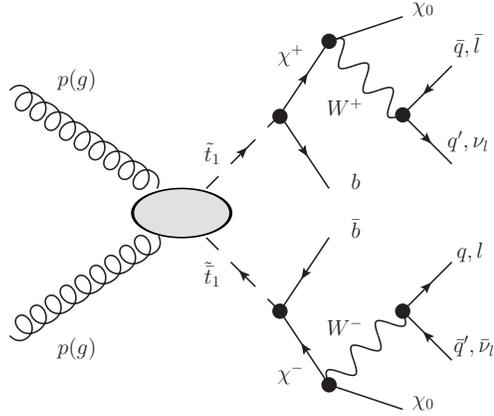} 
	\end{center}
	\caption{The lightest stop pair-production process at the LHC energies in proton-proton collisions. 
	The blob corresponds to all the tree-level diagrams contributing to the stop production.  
	The final states include two b-jets, missing energy 
$\Big/\hspace{-0.3cm E_T}$, the
	 light-quark jets, and leptons.
	With almost equal probability (45 \%) the virtual $W$-boson produces either four jets or two jets accompanied by 
	a charged lepton and additional missing energy from neutrino. In 10 \% of cases two $W$-bosons decay leptonically, 
	and instead of the light-quark jets we have two charged leptons and additional 
$\Big/\hspace{-0.3cm E_T}$
	from two neutrinos.}
	\label{fig:stop_prod}
\end{figure}

It turns out that the chargino decays into the light quarks with 66 \% probability. In 33 \% cases it produces leptons. 
As a consequence, we have the following key signature for the stop pair production:  two b-jets coming from the decay of the stops, missing energy 
$\Big/\hspace{-0.3cm E_T}$
from two neutralinos, and light-quark jets or leptons from the virtual $W$-bosons (see Fig.~\ref{fig:stop_prod}).
It is obvious that for the considered value of the stop mass the final states are similar 
to that of the top pair production so one can search for $\tilde t_1 \bar \tilde t_1$ signal in the $t\bar t$ event sample
as it was done in Ref.~\refcite{Abazov:2009ps}.

The ATLAS collaboration has already performed a study of such signatures\cite{Aad:2011ks} at $\sqrt {s} = 7$ TeV
with real data obtained in 2010 (so-called one-lepton analysis with b-jets and missing transverse energy). 
Their results can be interpreted as exclusion limits in the $(m_{\tilde g},\ms{t}{1}$) plane 
($m_{\tilde g}$ being the gluino mass) and, according to Fig.~3 of Ref.~\refcite{Aad:2011ks},  
the stop production cross-section should be smaller than 15-40 pb for $\ms{t}{1}\simeq 180$~GeV depending on the gluino mass 
which varies in the range 350-620~GeV. Since for our benchmark point the cross-section of stop pair production with the given final states is approximately 
$\sigma_{\tilde t_1 \bar \tilde t_1} \times \mathrm{BR}(\tilde t_1 \bar \tilde t_1 \to b q q' b  l \nu) = 8 \times 0.45 = 3.6$~pb 
and $m_{\tilde g} \simeq 630$ GeV, it seems that we escape the current ATLAS bound.  However,  the searches for the light stop production seems to be very challenging and we attract attention to this decay mode.

Another interesting point is that the $B_s\to \mu^+ \mu^-$ branching fraction almost touches the experimentally allowed boundary line so it may happen that this rare decay would be observed at the LHCb experiment in the near future.

Thus, our main conclusion is that there exists a possibility
when the cross-section of the single Higgs production is large enough to favour its observation at the LHC even with intermediate luminosity. In addition, the search for the lightest stop production in the $\tilde t \to \chi^+ b$ mode seems to be within the reach of the LHC at the early stage.
Whether we are lucky or not  will be clear only a posteriori.  However, any favourable possibility should not be missed.

\section*{Acknowledgements}

We are grateful to Wim de Boer, F. Ratnikov, and V.A. Bednyakov for useful discussions.  Financial support
from RFBR grant \# 11-02-01177 and the Ministry of Education and
Science of the Russian Federation grant \# 1027.2008.2   and the Heisenberg-Landau Programme is kindly
acknowledged.  D.K. and \c{S}.H.T. would like to thank Karlsruhe University for hospitality.


\begin{thebibliography}{99}


\bibitem{Djouadi:2005gj}
  A.~Djouadi,
  {\it Phys.\ Rep.}  {\bf 459},  1 (2008)  
  hep-ph/0503173.
  
\bibitem{Dittmaier:2011ti}
  LHC Higgs Cross Section Working Group Collaboration (S.~Dittmaier {\it et al.}), 
  Handbook of LHC Higgs Cross Sections: 1. Inclusive Observables,
  arXiv:1101.0593 [hep-ph].
  
\bibitem{Beskidt:2010va}
  C.~Beskidt, W.~de Boer, T.~Hanisch, E.~Ziebarth, V.~Zhukov, D.I.~Kazakov,
  {\it Phys.\ Lett. B } {\bf 695},  143 (2011),  
  arXiv:1008.2150 [hep-ph].
  
\bibitem{Georgi:1977gs}
  H.~M.~Georgi, S.~L.~Glashow, M.~E.~Machacek {\it et al.},
  {\it Phys.\ Rev.\ Lett.}  {\bf 40},  692 (1978).
 
 
\bibitem{Barberio:2008fa}
 Heavy Flavor Averaging Group (E.~Barberio {\it et al.}),
 Averages of $b-$hadron and $c-$hadron Properties at the End of 2007,
 arXiv:0808.1297 [hep-ex].
 
\bibitem{Bmumuexp}
  CDF Collaboration (T.~Aaltonen {\it et al.}),
  {\it Phys.\ Rev.\ Lett.}  {\bf 100}, 101802 (2008),
  arXiv:0712.1708 [hep-ex].

\bibitem{Bmumuexp2}
  D0 Collaboration (V.~M.~Abazov {\it et al.}),
  {\it Phys.\ Lett. B}  {\bf 693 },   539  (2010), 
  arXiv:1006.3469 [hep-ex].
  

\bibitem{g-2}
  D.~Stockinger,
  {\it J.\ Phys.\ G} {\bf 34}, R45 (2007), 
  arXiv:hep-ph/0609168]. 


\bibitem{Komatsu:2008hk}
  WMAP Collaboration (E.~Komatsu {\it et al.}),
  {\it Astrophys.\ J.\ Suppl.} {\bf 180 },  330  (2009),
  arXiv:0803.0547 [astro-ph].

\bibitem{Heinemeyer:2004gx}
  S.~Heinemeyer, W.~Hollik, G.~Weiglein,
  {\it Phys.\ Rept.\ } {\bf 425}, 265 (2006),
  hep-ph/0412214.


\bibitem{Ellis:2002wv}
  J.~R.~Ellis, K.~A.~Olive, Y.~Santoso,
  {\it Phys.\ Lett.\ B}  {\bf 539 }, 107 (2002),
  hep-ph/0204192.
  
\bibitem{Spira:1993bb}
 M.~Spira, A.~Djouadi, D.~Graudenz, P.M.~Zerwas,
 {\it Phys.\ Lett.\ B} {\bf 318 }, 347 (1993). 
  

\bibitem{haber}
J. F. Gunion, H. Haber, G. Kane and S. Dawson, \textit{The Higgs Hunter’s Guide}, (Addison-
Wesley Publishing Company, Redwood City, CA, 1990).

\bibitem{okun} 
L. B. Okun, \textit{Leptons and Quarks}, (Elsevier Science Pub Co, March 1, 1985).


\bibitem{Allanach:2001kg}
  B.~C.~Allanach,
  {\it Comput.\ Phys.\ Commun.\ }  {\bf 143}, 305 (2002),
  hep-ph/0104145.

\bibitem{Carena:1999py}
  M.~S.~Carena, D.~Garcia, U.~Nierste, C.~E.~M.~Wagner,
  {\it Nucl.\ Phys.\ B}  {\bf 577}, 88 (2000),
  hep-ph/9912516. 


\bibitem{PDFs1}
  A.~D.~Martin, W.~J.~Stirling, R.~S.~Thorne and G.~Watt,
  {\it Eur.\ Phys.\ J.\  C} {\bf 63}, 189 (2009),
  arXiv:0901.0002 [hep-ph].

\bibitem{PDFs2}

  A.~D.~Martin, W.~J.~Stirling, R.~S.~Thorne and G.~Watt,
  {\it Eur.\ Phys.\ J.\  C} {\bf 64},  653 (2009), 
  arXiv:0905.3531 [hep-ph].

\bibitem{PDFs3}
  A.~D.~Martin, W.~J.~Stirling, R.~S.~Thorne and G.~Watt,
  {\it Eur.\ Phys.\ J.\  C} {\bf 70}, 51  (2010),
  arXiv:1007.2624 [hep-ph].


 \bibitem{nnlo1} 
  R.~V.~Harlander, M.~Steinhauser,
  {it JHEP} {\bf 0409}, 066  (2004),
  hep-ph/0409010.
 \bibitem{nnlo2} 
  C.~Anastasiou, S.~Beerli, A.~Daleo,
  {\it Phys.\ Rev.\ Lett.\ } {\bf 100}, 241806 (2008),
  arXiv:0803.3065 [hep-ph].
  
 \bibitem{n3lo1} 
  R.~V.~Harlander, K.~J.~Ozeren,
  {\it JHEP} {\bf 0911}, 088 (2009),
  arXiv:0909.3420 [hep-ph].

 \bibitem{n3lo2} 
  A.~Pak, M.~Rogal, M.~Steinhauser,
  {\it JHEP} {\bf 1002}, 025 (2010),
  arXiv:0911.4662 [hep-ph].

 \bibitem{n3lo3} 
  R.~V.~Harlander, H.~Mantler, S.~Marzani {\it et al.},
  {\it Eur.\ Phys.\ J.\ C} {\bf 66}, 359 (2010),
  arXiv:0912.2104 [hep-ph].

 \bibitem{n3lo4} 
  A.~Pak, M.~Steinhauser, N.~Zerf,
  Higgs boson production in gluon fusion to NNLO in the MSSM,
  arXiv:1012.0639 [hep-ph].
  
\bibitem{Muhlleitner:2006wx}
  M.~Muhlleitner, M.~Spira,
  {\it Nucl.\ Phys.\ B} {\bf 790}, 1 (2008),
  hep-ph/0612254.
  
    
 \bibitem{Kaz}
  D.~I.~Kazakov,
  Supersymmetry on the Run: LHC and Dark Matter, in 
  {\it Nucl.\ Phys.\ Proc.\ Suppl.\ } {\bf 203-204}, 118 (2010),
  arXiv:1010.5419 [hep-ph].

  
\bibitem{CK}
  S.~Codoban and D.~I.~Kazakov,
  {\it Eur.\ Phys.\ J.\  C} {\bf 13}, 671 (2000),
  hep-ph/9906256.

\bibitem{Barbieri:1993av}
  R.~Barbieri, G.~F.~Giudice,
  {\it Phys.\ Lett.\ B}  {\bf 309}, 86 (1993),
  hep-ph/9303270.

\bibitem{Degrassi:2000qf}
  G.~Degrassi, P.~Gambino, G.~F.~Giudice,
  {\it JHEP} {\bf 0012}, 009 (2000),
  hep-ph/0009337.

\bibitem{superiso1}
  F.~Mahmoudi,
  {\it Comput.\ Phys.\ Commun.\ } {\bf 180}, 1579 (2009),
  arXiv:0808.3144 [hep-ph].
\bibitem{superiso2}
  F.~Mahmoudi,
  {\it Comput.\ Phys.\ Commun.\ } {\bf 180}, 1718 (2009)

\bibitem{superiso_relic}
  A.~Arbey, F.~Mahmoudi,
  {\it Comput.\ Phys.\ Commun.\ } {\bf 181}, 1277 (2010),
  arXiv:0906.0369 [hep-ph].

\bibitem{Hahn:2010te}
  T.~Hahn, S.~Heinemeyer, W.~Hollik, H.~Rzehak, G.~Weiglein,
  {\it Nucl.\ Phys.\ Proc.\ Suppl.\ }  {\bf 205-206}, 152 (2010),
  arXiv:1007.0956 [hep-ph].



\bibitem{Bobeth:2001sq}
  C.~Bobeth, T.~Ewerth, F.~Kruger, J.~Urban,
  {\it Phys.\ Rev.\ D} {\bf 64}, 074014 (2001),
  hep-ph/0104284.

\bibitem{Schumacher:2011jq}
ATLAS Collaboration (M.~Schumacher),
Higgs Boson Searches with ATLAS based on 2010 Data,
	      arXiv:1106.2496 [hep-ex].

\bibitem{Chatrchyan:2011nx}
CMS Collaboration (S.~Chatrchyan {\it et al.}),
Search for Neutral MSSM Higgs Bosons Decaying to Tau Pairs in $pp$ Collisions at $\sqrt{s}=7$ TeV,
	      ,arXiv:1104.1619 [hep-ex].


  
\bibitem{Bechtle:2011sb}
  P.~Bechtle, O.~Brein, S.~Heinemeyer, G.~Weiglein and K.~E.~Williams,
  HiggsBounds 2.0.0: Confronting Neutral and Charged Higgs Sector Predictions
  with Exclusion Bounds from LEP and the Tevatron,
  arXiv:1102.1898 [hep-ph].

\bibitem{Pukhov:2004ca}
  A.~Pukhov,
  Calchep 2.3: MSSM, structure functions, event generation, and generation
  of matrix elements for other packages,
  arXiv:hep-ph/0412191.

\bibitem{Harlander:2003ai}
  R.~V.~Harlander and W.~B.~Kilgore,
  {\it Phys.\ Rev.\  D} {\bf 68}, 013001 (2003),
  arXiv:hep-ph/0304035.

\bibitem{Djouadi:2006bz}
  A.~Djouadi, M.~M.~Muhlleitner, M.~Spira,
  {\it Acta Phys.\ Polon.\ B}  {\bf 38}, 635 (2007),
  hep-ph/0609292.

  
  
\bibitem{Abazov:2009ps}
  DO Collaboration (V.~M.~Abazov {\it et al.}),
  {\it Phys.\ Lett.\ B}  {\bf 674}, 4 (2009).
  arXiv:0901.1063 [hep-ex].
  

\bibitem{Aad:2011ks}
  ATLAS Collaboration (G.~Aad {\it et al.}),
  Search for supersymmetry in pp collisions at $\sqrt{s} = 7$ TeV in final states with missing transverse momentum and b-jets,
 arXiv:1103.4344 [hep-ex].  


\end{thebibliography}
\end{document}